\documentclass[runningheads]{llncs}
\usepackage[T1]{fontenc}
\usepackage{graphicx}
\usepackage{textcomp}
\usepackage{xcolor}
\usepackage{tabularray}
\usepackage{tcolorbox}
\usepackage[hyphens]{url}
\usepackage{hyperref}
\usepackage{multirow}
\usepackage{lscape}

\usepackage[margin=1in]{geometry}

\usepackage[normalem]{ulem}
\useunder{\uline}{\ul}{}
\usepackage{booktabs}
\usepackage{amsmath}
\usepackage{amsmath}
\usepackage{amstext}
\usepackage{tcolorbox}
\newtcolorbox{takeawaybox}[1][]{
  colframe=gray!75,      
  colback=white!10,       
  coltitle=black,        
  boxrule=0.5mm,         
  fonttitle=\bfseries,   
  title=#1               
}
\hypersetup
{
  colorlinks   = true, 
  urlcolor     = blue, 
  linkcolor    = blue, 
  citecolor    = blue 
}
\usepackage{tabularx}
\usepackage{array}
\usepackage{booktabs}

\usepackage[utf8]{inputenc}

\begin{document}
\title{LLM-based Multi-Agent System for Intelligent Refactoring of Haskell Code}

\author{Anonymous Author(s)}

\authorrunning{Anonymous et al.}
\titlerunning{LLM-based Multi-agent Refactoring of Haskell Code}

%
%


\author{Shahbaz Siddeeq\inst{1} \and
Muhammad Waseem\inst{1} \and
Zeeshan Rasheed\inst{1} \and
Md Mahade Hasan\inst{1} \and
Jussi Rasku\inst{1} \and
Mika Saari\inst{1} \and
Henri Terho\inst{2} \and
Kalle Mäkelä\inst{2} \and
Kai-Kristian Kemell\inst{1} \and
Pekka Abrahamsson\inst{1}}
%

\authorrunning{S. Siddeeq et al.}

%

%
\institute{Faculty of Information Technology and Communication Sciences, Tampere University, Finland \\
\email{\{shahbaz.siddeeq, muhammad.waseem, zeeshan.rasheed, mdmahade.hasan, jussi.rasku, mika.saari, Kai-Kristian.Kemell, pekka.abrahamsson\}@tuni.fi} \and
Eficode Oy, Finland \\
\email{\{henri.terho\}@eficode.com, \{kalle.makela\}@gmail.com}}

\maketitle              
\begin{abstract}

Refactoring is a constant activity in software development and maintenance. Scale and maintain software systems are based on code refactoring. However, this process is still labor intensive, as it requires programmers to analyze the codebases in detail to avoid introducing new defects. In this research, we put forward a large language model (LLM)-based multi-agent system to automate the refactoring process on Haskell code. The objective of this research is to evaluate the effect of LLM-based agents in performing structured and semantically accurate refactoring on Haskell code. Our proposed multi-agent system based on specialized agents with distinct roles, including code analysis, refactoring execution, verification, and debugging. To test the effectiveness and practical applicability of the multi-agent system, we conducted evaluations using different  open-source Haskell codebases. The results of the experiments carried out showed that the proposed LLM-based multi-agent system could average 11.03\% decreased complexity in code, an improvement of 22.46\% in overall code quality, and increase performance efficiency by an average of 13.27\%. Furthermore, memory allocation was optimized by up to 14.57\%. These results highlight the ability of LLM-based multi-agent in managing refactoring tasks  targeted toward functional programming paradigms. Our findings hint that LLM-based multi-agent systems integration into the refactoring of functional programming languages can enhance maintainability and support automated development workflows.

\end{abstract}

\section{Introduction} \label{sec:intro}
Functional programming languages are known for their higher-order functions and immutability, which make them suitable for accurate and concurrent processing \cite{hu2015functional}. This core strength of functional programming is showcased in languages like Haskell \cite{hudak1992gentle}, which provide various features designed for such tasks, including type inference, lazy evaluation, and monadic handling of side effects~\cite{peyton1993imperative}. However, these characteristics also make Haskell based systems challenging to refactor, as modifications must account for dependencies and constructs like monads \cite{figueroa2021monads} and type classes \cite{orchard2014embedding}.

Refactoring functional programming languages like Haskell based applications presents a set of challenges~\cite{brown2011expression}. Haskell's functional nature introduces complexity in maintaining code readability, structure, and performance without compromising immutability and higher-level abstractions~\cite{mens2004survey} unlike imperative languages. Traditionally, refactoring efforts in functional languages focus on restructuring code to improve readability and maintainability while ensuring correctness, but the presence of type systems and non-linear evaluation strategies adds difficulty to the process.

Previous research has explored the use of multi-agent systems  in software engineering for tasks such as automated analysis \cite{wooldridge1995intelligent}, distributed processing \cite{abdallah2011dynamic}. These studies have primarily focused on general-purpose programming languages and conventional paradigms. Similarly, the adoption of large language models (LLMs) for code generation, error detection, and basic refactoring has demonstrated potential, but their application to functional programming languages like Haskell remains limited. Existing tools for Haskell refactoring, such as HaRe~\cite{li2005haskell} rely on static or rule-based approaches and do not use the advanced capabilities of LLMs or the collaborative nature of multi-agent systems. Multi-agent systems interact, collaborate, and execute tasks concurrently, which is making them a system for modular and distributed processes \cite{abdallah2011dynamic}. In software maintenance, agents have been applied to distributed code analysis and task allocation, with each agent specializing in a function, thereby streamlining processes. Tasks like code analysis, code refactoring, and verification can be distributed with agents to address the structured manner complexity by applying multi-agent system principles to Haskell code refactoring.

Haskell based application poses challenges while refactoring due to features like type safety, immutability, and lazy evaluation. These characteristics demand efficiency in code refactoring while maintaining its functional integrity. Traditional systems often rely on rule-based systems or high manual resource interventions such as HaRe~\cite{li2005haskell}, which lack the adaptability and scalability required for large and complex codebases. LLMs have shown promise in understanding and generating code, yet they struggle with the specific demands of functional programming~\cite{shirafuji2023refactoring,jiang2024survey}. This research aims to overcome these limitations by integrating LLM-based multi-agent system to distribute refactoring tasks.

\textbf{Motivation}: The motivation behind this research is to meet an industry demand for automated refactoring system for complex codebases. LLMs like Codex \cite{shirafuji2023refactoring} and GPT \cite{ishizue2024improved} have transformed the programming landscape, offering capabilities in error detection, code synthesis, and refactoring. LLMs have shown promise in understanding and generating code structures, facilitating tasks such as bug detection~\cite{chen2021evaluating,svyatkovskiy2020intellicode} and automatic code completion by training on code datasets. Recent studies have started to explore the abilities of LLMs in refactoring tasks~\cite{pomian2024next,dos2015autorefactoring}, where their ability to analyze context and suggest changes has highlights their effectiveness in code enhancement. However, despite these advances, LLMs alone are often insufficient for managing the requirements of functional programming languages, especially in distributed or modular tasks, where multi-agent system approaches may provide support~\cite{guo2024large}.

\textbf{Objective of the Study:} The objective of this study is to develop LLM based multi-agent system to facilitate the refactoring of Haskell codebases. The system aims to achieve the following: (i) improve code quality metrics, specifically cyclomatic complexity, runtime efficiency, memory usage, and HLint comparison, and (ii) ensure that the system is adaptable to the characteristics of functional programming paradigms such as lazy evaluation, immutability, and type safety. The primary \textbf{contributions} of this study are: 
\begin{itemize}
    \item Development of an LLM-based multi-agent system for refactoring Haskell code.
    \item Evaluation of the developed system against code quality metrics, including cyclomatic complexity, runtime efficiency, memory usage, and comparisons with HLint.
     \item Public release of the developed system, which is made available online~\cite{ShahbazMaSHaskellGitHub2025}, allowing researchers and practitioners to access, replicate, and validate the system. 
\end{itemize}

\textbf{Paper Organization}: Section \ref{sec:rw} describes the related work; Section \ref{sec:rm} presents the experimental design; Section \ref{sec:results} describes the results and evaluation; Section \ref{sec:discussion} discusses this work with limitations; Section \ref{sec:threat} describes threats to validity; Section \ref{sec:conclusions} concludes this work and future research directions.

\section{Related Work} \label{sec:rw}

In this section, we review studies relevant to our research. Section \ref{sec:rw1} provides an overview of functional programming languages and Haskell refactoring. In Section \ref{sec:rw3}, we discuss the use of LLMs for code generation and refactoring, as well as the integration of LLM-based multi-agent systems for code refactoring.

\subsection{Functional Programming Languages and Haskell Refactoring}\label{sec:rw1}
Functional programming languages like Haskell present challenges due to their emphasis on immutability, lazy evaluation, and higher-order functions~\cite{bragilevsky2021haskell}. These
challenges complicate the refactoring process, especially when changes need to preserve program semantics~\cite{thompson2013refactoring}. Most research on refactoring functional programming-based applications primarily focuses on traditional methods. Mens and Tourwé~\cite{mens2004survey} conducted a comprehensive survey highlighting refactoring patterns applicable across programming paradigms, including managing side effects and optimizing type safety. In addition, the challenges of refactoring functional programs have been explored, with techniques proposed that are specifically designed for functional codebases to maintain semantic integrity while improving readability and performance ~\cite{abdallah2011dynamic}.

Despite efforts, there is still limited research on automated refactoring in Haskell and most methods depending on rule-based systems. These limitations highlight the need for automated systems that can understand dependencies within Haskell code and suggest effective refactoring solutions. Our research seeks to bridge this gap by introducing a multi-agent system capable of performing distributed and automated refactoring tasks within a Haskell codebase.

\subsection{LLMs for Code Generation and Refactoring}\label{sec:rw3}
The rise of LLMs has revolutionized automated code generation and refactoring such as OpenAI’s GPT series \cite{ishizue2024improved} and Codex \cite{chen2021evaluating}. These models, trained on vast datasets of code, have demonstrated proficiency in tasks ranging from code completion to error detection and refactoring suggestions~\cite{chen2021evaluating}. Svyatkovskiy \textit{et al}.~\cite{svyatkovskiy2020intellicode} introduced IntelliCode Compose reducing the time developers spend on routine coding tasks which is a transformer-based system that uses LLMs to assist in code generation. Recent work~\cite{white2024chatgpt} has shown that LLMs can not only generate syntactically correct code but also adapt to specific programming paradigms, making them valuable assets for software refactoring. The LLMs based multi-agent systems for code refactoring is an approach that has not been extensively explored in the literature. Recently Baumgartner \textit{et al}.~\cite{baumgartner2024ai} a multi-agent learning system for automatic code refactoring, which demonstrated improvements in handling distributed code modifications. However, their approach was primarily focused on imperative programming languages and did not explore applications in functional programming. Similarly, discussed~\cite{huang2024levels} a multi-agent approach to software maintenance but their work did not consider the challenges associated with functional languages like Haskell.

Our research expands on these prior works by introducing a multi-agent system~\cite{hua2023war} specifically designed for functional programming refactoring~\cite{dos2015autorefactoring}, enhanced by the contextual analysis and code generation capabilities of LLMs. By assigning each agent a specialized task and utilizing LLMs for deeper language comprehension, our approach aims to simplify Haskell refactoring ~\cite{thompson2013refactoring} in a scalable, autonomous manner. This combination of multi-agent systems and LLMs provides a solution to the challenges of functional programming language maintenance and improves the utility of both LLMs and multi-agent systems in software engineering~\cite{cheng2024exploring}.

\section{Experimental Design} \label{sec:rm}
\begin{figure*}[h]
    \centering
    \includegraphics[width=0.99\textwidth, clip, trim=20 50 100 40]{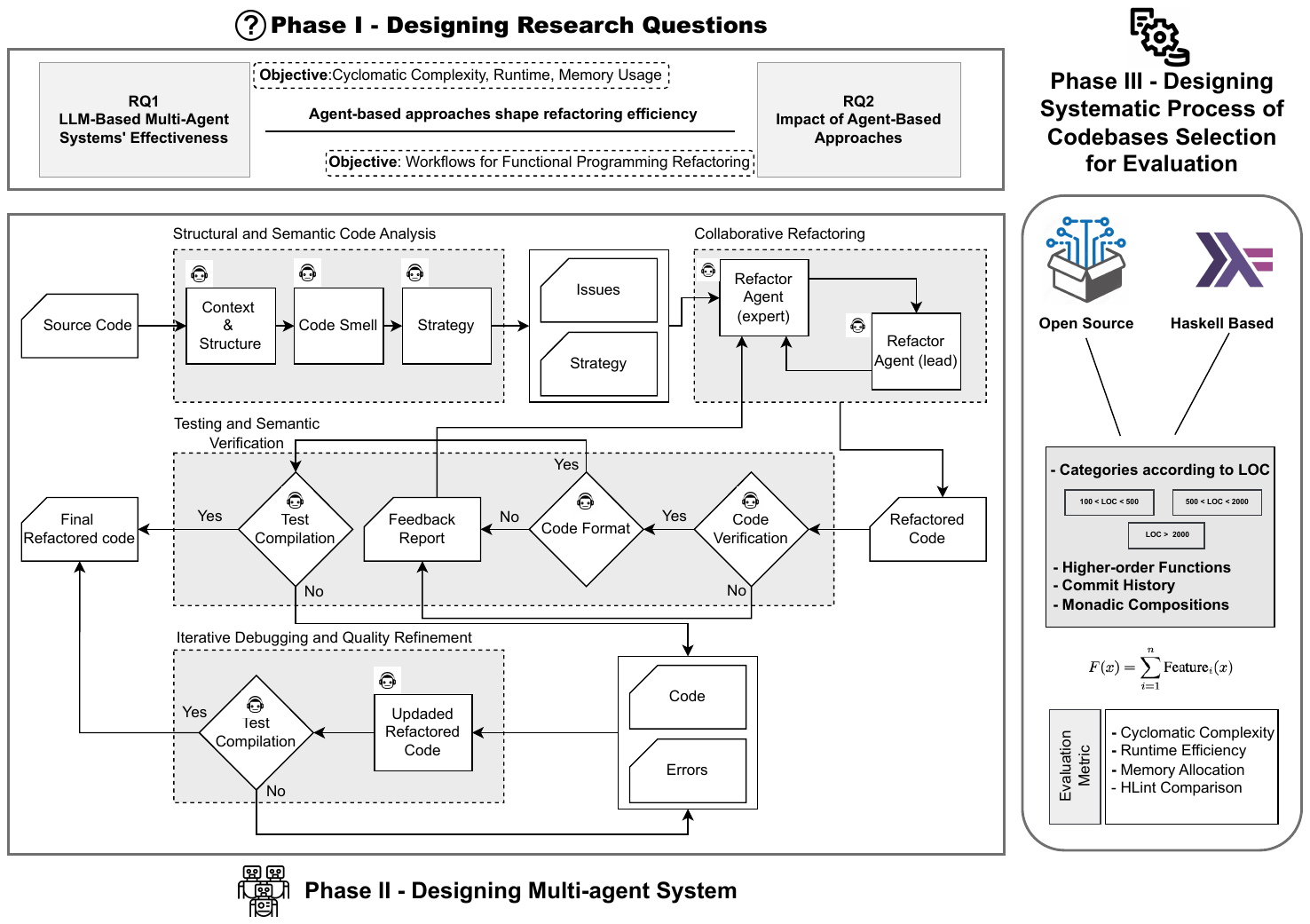}
    \caption{Overview of the experimental design illustrating the three main phases: (I) Designing research questions focused on evaluating the LLM effectiveness and agent-based approaches impact; (II) Designing a multi-agent system architecture for structural and semantic code analysis, collaborative refactoring, and iterative debugging; and (III) Designing the systematic process of open-source and Haskell-based codebases selection for evaluation.}
    \label{fig:RM}
\end{figure*}

The experimental design for this study is divided into three phases, which are elaborated below and illustrated in Fig. \ref{fig:RM}.

\subsection{Phase I - Designing Research Questions}
Considering our research objective, we formulate the following two RQs.

\begin{tcolorbox}[colback=gray!2!white,colframe=black!75!black]
\textit{\textbf{RQ1.} How effectively can LLM-based multi-agent systems improve Haskell codebases in terms of cyclomatic complexity, runtime, and memory usage?}
\end{tcolorbox}

The \textbf{objective} of this RQ is to evaluate the effectiveness of LLM-based multi-agent systems in automating Haskell code refactoring, specifically focusing on reducing cyclomatic complexity, improving runtime performance, and optimizing memory usage.

\begin{tcolorbox}[colback=gray!2!white,colframe=black!75!black]
\textit{\textbf{RQ2.} What is the impact of using multi-agent approaches on refactoring workflows for functional programming languages?}
\end{tcolorbox}
      
The \textbf{objective} of this RQ is to assess the flexibility of a multi-agent system in addressing the challenges of refactoring Haskell code, particularly focusing on aspects such as immutability, lazy evaluation, and type safety.

\subsection{Phase II- Designing Multi-agent System}
In this phase, we focus on the design and management of our multi-agent system, beginning with the development of specialized prompt strategies for each agent and culminating in a collaborative architecture designed for intelligent Haskell code refactoring. This outlines how well-defined agent roles and structured communication workflows enable effective and context-aware refactoring within functional programming environments. The system is model agnostic in practice (can be implemented with any LLM), but that we have tested it specifically with GPT-4o.

\subsubsection{Prompt Strategies}
\begin{figure}[h]
\centering
\includegraphics[width=0.99\textwidth, clip, trim=10 560 10 30]{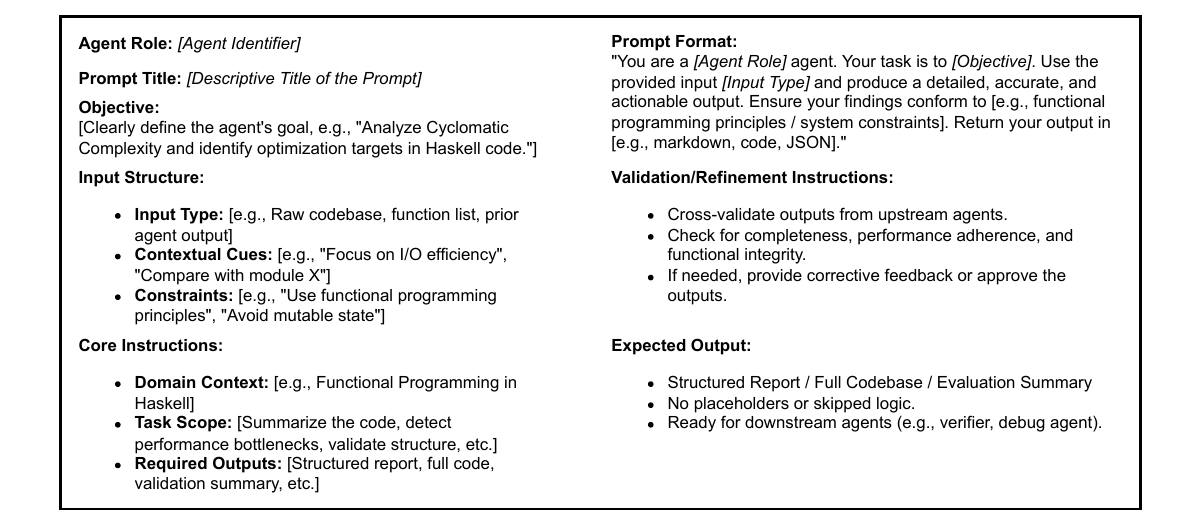}
\caption{Structured Prompt Template for Multi-agent System Roles}
\label{fig:agent_prompt_strucutre}
\end{figure}

The success of a multi-agent system relies on the precision and adaptability of its prompt engineering. In this work, prompts are designed to ensure role-specific task delegation, facilitate inter-agent collaboration, and follow the principles of functional programming. Each agent operates on detailed, structured prompts tailored to its specific role within the multi-agent system and highlighted the structure in Fig.~\ref{fig:agent_prompt_strucutre}. The prompt strategies in this study are informed by the following:

\begin{itemize}
    \item \textbf{Role-specific Prompting}: Tailored prompts enable agents to focus on specific tasks. This approach builds on prior research highlighting task-specific prompt engineering in LLMs~\cite{brown2020language,feng2024genetic}.
    \item \textbf{Dynamic and Iterative Prompt Refinement}: Prompts are updated based on intermediate outputs, aligning with strategies employed in adaptive language models~\cite{nascimento2023self}.
\end{itemize}
Our approach used an agentic framework where each agent handles part of the refactoring process rather than running several instances of an LLM independently. Key agents and their prompts include:

\begin{itemize}
   \item \textbf{Code Refactor Agent}: Prompts are designed to extract the purpose of the codebase, identify module dependencies, and summarize function interrelationships. This strategy uses methods similar to structured natural language processing for software engineering tasks~\cite{allamanis2018survey}.
   \item \textbf{Verification Agent}: Employ prompts that emphasize validation, ensuring outputs from context and analysis agents meet the required functional and structural criteria.
   \item \textbf{Code Debug Agent}: Focused on inefficiencies and performance bottlenecks, its prompts integrate principles from functional programming analyses, as outlined in Gill~\cite{gill2009worker} and Wadler~\cite{wadler1992essence}. 
\end{itemize}

\begin{tcolorbox}[colback=gray!2!white, colframe=gray]
\textbf{Hybrid Prompt Type:}
Every agent prompt begins by specifying the agent’s role \textbf{(role-based)}, gives explicit task instructions \textbf{(instructional)}, embeds relevant context such as code or previous agent outputs \textbf{(contextual)}, and, where applicable, directs the agent to check or critique outputs for correctness or adherence to best practices \textbf{(validation)}. This hybrid prompt design ensures each agent acts with both autonomy and accountability, supporting robust and coordinated multi-agent collaboration throughout the refactoring pipeline.

\end{tcolorbox}

These strategies ensure communication between agents and improve task-specific outputs, thus enhancing the overall effectiveness of the multi-agent system.

\subsubsection{Multi-agent System}
In this phase, we present a LLM-based multi-agent system that performs automated refactoring on Haskell code.The aim is to use of agents that collaboratively identify code issues, improving the structure and quality of the code. Each agent within this multi-agent system perform independently but remains coordinated to confirm consistent improvement in the codebase. Our multi-agent system is designed so that each agent performs  its specialized task autonomously with its dedicated logic and prompt structure for tasks such as  code smell detection, refactoring, verification, or debugging. However, these agents  work together in a coordinated workflow, where the output of one agent becomes the input or context for the next agent rather than working in isolation. This inter-agent communication is managed via structured data exchanges and prompt chaining, ensuring that all agents are aligned toward the shared objective of improving the codebase. The system achieves iterative code refinement.

\begin{itemize}

    \item \textbf{Structural and Semantic Code Analysis:}
    The process starts with an analysis of the source code. The analysis focuses on components of functional programming, such as monadic compositions and higher-order functions which impact on code maintainability. The multi-agent system then undertakes a detailed contextual and structural analysis (e.g., Layered Architecture \cite{rich1985layered}, Hexagonal \cite{griffin2021hexagonal,he2005hexagonal}, Onion architecture \cite{khalil2016onion}) of the code segments. This involves analyzing the source code to find code smells, usage trends, and architectural features. This analysis provides a strong basis for the refactoring strategies that are subsequently developed to address particular inefficiencies.
    
    \begin{itemize}
        \item \textbf{Code Context and Structure Agent:} The Code Context and Structure Agent is responsible for analyzing the codebase and providing a clear structural overview. It begins by parsing the code to identify key components, such as modules, functions, and their interdependencies. These identifications provide essential context for subsequent agents, ensuring that their analysis and refactoring tasks are precise and effective.
        \item \textbf{Code Smells Agent:} We grouped the 42 refactoring items into eight high-level categories, then merged them into agent responsibilities which is shown in Table \ref{tab:refactor_checklists}. Agent conducts a thorough evaluation of the codebase to identify inefficiencies and areas requiring improvement. It calculates key metrics, such as Cyclomatic Complexity, using McCabe’s formula: \textit{CC = E - N + 2P} where \textbf{\textit{E}} represents the number of edges in the control flow graph, \textbf{\textit{N}} is the number of nodes, and \textbf{\textit{P}} denotes the number of connected components or entry points. The agent identifies areas with high complexity and low modularity, marking them for refactoring. This analysis helps create refactoring strategies that improve structure and maintainability.

        \begin{table*}[ht!]
        \caption{High-level categories for Haskell refactoring checklist}
        \centering
        \label{tab:refactor_checklists}
        \renewcommand{\arraystretch}{1.2}
        \resizebox{\textwidth}{!}{%
        \newcolumntype{L}[1]{>{\hspace{#1}}l}
        \begin{tabular}{|L{8pt}|L{8pt}|L{8pt}|L{8pt}|}
        \hline
        \textbf{Category}                        & \textbf{Checklist}                                                                                                                                                                                                                                                                                                                                                                                                                           & \textbf{Category}                 & \textbf{Checklist}                                                                                                                                                                                                                                                                                                                                                                                                                                                                          \\ \hline
        \begin{tabular}[c]{@{}l@{}}1. Code Cleanup \\ \& Simplification\end{tabular}  & \begin{tabular}[c]{@{}l@{}}1. Remove unused imports\\ 2. Remove dead code (unused functions or values)\\ 3. Eliminate redundant let or where bindings\\ 4. Remove duplicate or repeated code\\ 5. Simplify boolean expressions and conditions\\ 6. Remove unnecessary parentheses and lambdas\\ 7. Collapse nested case or if statements\end{tabular}                                                                               & \begin{tabular}[c]{@{}l@{}}5. Safety \\ \& Robustness \end{tabular}    & \begin{tabular}[c]{@{}l@{}}22. Prefer pure functions and minimize side effects\\ 23. Use monads and applicatives appropriately\\ 24. Avoid partial functions like head, tail, fromJust\\ 25. Handle all data constructors explicitly in pattern matches\\ 26. Use strictness annotations (BangPatterns, seq) to avoid space leaks\end{tabular}                                                                                                                                     \\ \hline
        \begin{tabular}[c]{@{}l@{}}2. Readability \\ \& Style\end{tabular}            & \begin{tabular}[c]{@{}l@{}}8. Use descriptive function and variable names\\ 9. Add explicit type signatures to all top-level functions\\ 10. Add type annotations where inference harms readability\\ 11. Split large functions into smaller helpers\\ 12. Use where or let clauses meaningfully\\ 13. Use point-free style where it improves readability\\ 14. Avoid overusing point-free style when it hurts clarity\end{tabular} & \begin{tabular}[c]{@{}l@{}}6. Performance \\ Optimization\end{tabular} & \begin{tabular}[c]{@{}l@{}}27. Use foldl' for strict folding over large data\\ 28. Avoid creating unnecessary intermediate data structures\\ 29. Consider streaming libraries (conduit, pipes, etc.) for large data\\ 30. Replace primitive types with newtype or data for stronger type safety\\ 31. Benchmark and profile critical sections after refactoring\end{tabular}                                                                                                       \\ \hline
        \begin{tabular}[c]{@{}l@{}}3. Module \\ \& Dependency \\ Management\end{tabular} & \begin{tabular}[c]{@{}l@{}}15. Organize code into coherent modules\\ 16. Follow the single-responsibility principle in modules\\ 17. Avoid circular module dependencies\\ 18. Export only what's necessary from modules\\ 19. Hide internal implementations in module exports\end{tabular}                                                                                                                                          & 7. \begin{tabular}[c]{@{}l@{}}Functional \\ Idiom\end{tabular}         & \begin{tabular}[c]{@{}l@{}}32. Prefer map, foldr, foldl', zipWith, etc., over manual recursion\\ 33. Choose between list comprehensions and monadic code based on readability\\ 34. Use pattern matching instead of if-then-else where appropriate\\ 35. Use applicative style for independent computations\\ 36. Use maybe, either, or pattern guards to handle optional/alternative flows\\ 37. Avoid String in performance-sensitive code (use Text or ByteString)\end{tabular} \\ \hline
        \begin{tabular}[c]{@{}l@{}}4. Abstraction \\ \& Reusability\end{tabular}      & \begin{tabular}[c]{@{}l@{}}20. Abstract common logic using higher-order functions\\ 21. Define reusable patterns with typeclasses or pattern synonyms\end{tabular}                                                                                                                                                                                                                                                                  & 8. \begin{tabular}[c]{@{}l@{}}Continuous \\ Improvement\end{tabular}   & \begin{tabular}[c]{@{}l@{}}38. Use NonEmpty lists where empty lists are invalid\\ 39. Format code consistently (indentation, spacing, alignment)\\ 40. Use tools like hlint, weeder, and ghcid for continuous feedback\\ 41. Use literate Haskell (.lhs) where inline documentation is beneficial\\ 42. Use compiler warnings (-Wall, -Werror) to catch issues early\end{tabular}                                                                                                  \\ \hline
        \end{tabular}%
        }
        \end{table*}

        \item \textbf{Refactoring Strategy Agent:} The Refactoring Strategy Agent formulates refactoring strategies based on the analysis provided by the code smells Agent. It recommends modular decompositions, function simplifications, and optimizations of data structure to improve performance. This agent ensures that all strategies are customized to the characteristics of the codebase.
    \end{itemize}

    \item \textbf{Collaborative Refactoring through Multi-Agent Interaction:}
    Following the initial analysis, the system identifies code smells, patterns that suggest a weak code structure or possible issues. Each issue is documented, and refactoring strategies are defined to address them. specialized refactoring agents then execute these strategies. While each agent operates independently, their actions are coordinated to maintain consistency and resolve any potential conflicts across the codebase. These agents collaboratively work to enhance cyclomatic complexity, optimize runtime performance, and reduce memory allocation. Refactor Agents are responsible for implementing refactoring strategies while preserving the correctness and functionality of the codebase.
    \begin{itemize}
        \item \textbf{Refactor Agent (Expert):} Initiates the refactoring process. It focuses on removing redundancies, simplifying complex logic, and improving code readability. The refactor agent (expert) agent uses plain and simple refactoring rules, such as large functions breakdown,  duplicate code consolidation, and renaming variables for clarity. The refactor agent (expert) prepares the groundwork for later phases by improving a cleaner code structure.
        \item \textbf{Refactor Agent (Lead):} Building on the groundwork laid by the refactor agent (expert), refactor agent (lead) performs complex refactoring tasks. It focuses on  enforcing coding standards and optimizing performance. This agent refines the code further by addressing inefficiencies, reorganizing modules. The refactor agent (lead) refactors the code into a new structural strong form.
    \end{itemize}
    
    This collaborative model not only accelerates the refactoring process but also ensures a high degree of precision and adaptability to the unique characteristics of functional programming paradigms.

    \item \textbf{Testing and Semantic Verification:} Once the refactoring actions complete, the multi-agent system proceeds to a verification phase. Verification agents perform a series of automated tests to evaluate the  syntactic integrity, functional correctness, and compliance of the refactored code with quality standards. The feedback from these tests is crucial as it shows both successful improvements and areas that may require further intervention.
    \begin{itemize}
        \item \textbf{Testing and Validation Agent:} The Testing and validation agent tests the refactored code to ensure that it fulfills all functional and performance requirements. This agent confirms that the code is stable and ready for deployment by running test suites, including integration, unit, and system tests. Validation reports summarize the test results and highlight issues.
    \end{itemize}
    Verification agents analyze code style and formatting to follow Haskell practices. This testing ensures refactoring improves performance and keeps or improves readability and maintainability.

    \item \textbf{Iterative Debugging and Quality Refinement:} The system starts a focused cycle of debugging and improvement in response to the verification feedback. This iterative process involves re-executing selected refactoring actions, and re-testing the outcomes. This continues until all identified problems are fixed and the refactored code meets all non-functional and functional requirements.
    \begin{itemize}
        \item \textbf{Debug Agent:} The Debug Agent identifies and fixes problems that arise during the refactoring process. It makes sure the code runs as intended by analyzing runtime errors, and examining control flows. This agent is essential to maintain the reliability of the refactored codebase.
    \end{itemize}
    The iterative nature of this phase handles edge cases and unexpected issues, improving the strength and reliability of the final output.

    \item \textbf{Finalization:}
    The system creates documentation of strategies used, changes made, and improvements found. This helps with future maintenance, evaluation, and reuse of the method.
\end{itemize}
 This multi-agent system improves refactoring by meeting research goals and boosting performance and quality of functional programming code.

\subsection{Phase III - Designing Systematic Process of Codebases Selection for Evaluation}
This phase evaluates the effectiveness of multi-agent system using Open Source Projects (OSP).

\begin{itemize}
    \item \textbf{Deciding Codebases: }We included OSP to meet the goal of testing the multi-agent system under natural conditions. OSP provide real-world variability and development complexity. This ensures that our system is applicable to a broad spectrum of software projects.

    We chose Haskell codebases for evaluation based on a combination of quantitative and qualitative criteria. Quantitatively, the size of the codebase, measured in LOC, is a key parameter. Codebases are grouped into three categories: small ($100 \leq \text{LOC} < 500$), medium ($500 \leq \text{LOC} < 2000$), and large ($\text{LOC} \geq 2000$). This classification ensures that the system's scalability and performance are tested across various complexities. Additionally, the functional programming constructs present in the codebase, such as higher-order functions, type classes, and monadic compositions, are evaluated using a feature count metric:
    
    \begin{equation}
    F(x) = \sum_{i=1}^{n} \text{Feature}_i(x)
    \end{equation}

    where $\text{Feature}_i$ represents individual functional constructs. The inclusion of application domains, such as web development, data processing, and compilers, adds strength to the evaluation.

    \item \textbf{Selecting Codebases:} The OSP was curated through a systematic selection process presented in Table \ref{tab:systmatic_process} applied to public Haskell repositories on GitHub. Initially, repositories created before 2023, with more than 500 stars and written in Haskell, were shortlisted (213 total). Additional filters were applied to ensure activeness (updated within 12 months), collaborative development ($\geq$3 contributors), and adequate versioning complexity ($\geq$2 branches), narrowing the list to 133.

    \begin{table*}[ht!]
        \caption{Systematic repository selection process}
        \label{tab:systmatic_process}
        \centering
        \renewcommand{\arraystretch}{1}
        \resizebox{\textwidth}{!}{%
        \begin{tabular}{|>{\hspace{4pt}}l| >{\hspace{4pt}}l | >{\hspace{4pt}}l | >{\hspace{4pt}}l | >{\hspace{4pt}}l|}

            \hline
            \textbf{Step}                     & \textbf{Action}                                                                                                                                                                                                   & \textbf{Criteria / Rationale}                                                                                                                                                   & \textbf{Tools / Methods}                            & \textbf{Outcome}                                                            \\ \hline
            1. Initial Search                 & \begin{tabular}[c]{@{}l@{}}GitHub Advanced Search: \\ - created:\textless{}2023-01-01 \\ - stars:\textgreater{}500 \\ - language:Haskell\end{tabular}                                                               & \begin{tabular}[c]{@{}l@{}}- Maturity: \\ \hspace*{1em}Projects created before 2023. \\ - Popularity:\\ \hspace*{1em}\textgreater{}500 stars (community validation). \\ - Language: Haskell-focused.\end{tabular}    & \begin{tabular}[c]{@{}l@{}}GitHub \\ search syntax. \end{tabular}                              & 213 repos                                                            \\ \hline
            2. Meta-Filtering                 & \begin{tabular}[c]{@{}l@{}} - Remove inactive repos. (no updates in 12 months). \\ - Exclude solo projects (>2 contributors). \\ - Exclude repositories with \textless{}2 branches.\end{tabular} & \begin{tabular}[c]{@{}l@{}}- Activity: Ensure ongoing maintenance. \\ - Collaboration: Team-driven development. \\ - Complexity: Versioning/experimentation.\end{tabular}             & \begin{tabular}[c]{@{}l@{}}GitHub API \\(contributors, branches).\\ Manual review.\end{tabular} & 133 repos                                                            \\ \hline
            \begin{tabular}[c]{@{}l@{}}3. Commit \& \\ Documentation \\Filter\end{tabular} & \begin{tabular}[c]{@{}l@{}}- Scan commits for keywords: \\ \hspace*{1em}refactor, rewrite, cleanup, enhance, optimize\\ - Require docs with keywords: \\ \hspace*{1em}usage, install, example.\end{tabular}                                        & \begin{tabular}[c]{@{}l@{}}- Code Changes:\\ \hspace*{1em}Identify meaningful evolution patterns. \\ - Usability:\\ \hspace*{1em}Ensure documentation for analysis.\end{tabular}                                  & \begin{tabular}[c]{@{}l@{}}GitHub API scripted \\ keyword scans (regex).\end{tabular}          & 112 repos                                                            \\ \hline
            \begin{tabular}[c]{@{}l@{}}4. Size \\Categorization\end{tabular}            & \begin{tabular}[c]{@{}l@{}}Categorize by Lines of Code (LOC): \\ - Small: \textless{}500 LOC\\ - Medium: 500–2000 LOC\\ - Large: \textgreater{}2000 LOC\end{tabular}                                                    & \begin{tabular}[c]{@{}l@{}}- Diversity:\\ \hspace*{1em}Capture niche, mid-sized, and large. \\ - Scale:\\ \hspace*{1em}Avoid bias toward large codebases.\end{tabular}                                   & \begin{tabular}[c]{@{}l@{}}LOC analysis tools\\ (e.g., cloc).\\ GitHub Linguist.\end{tabular}   & \begin{tabular}[c]{@{}l@{}}Large: 106,\\ Medium: 4,\\ Small: 2\end{tabular} \\ \hline
            \begin{tabular}[c]{@{}l@{}}5. Proportional \\ Final Selection\end{tabular}   & Select 4 large, 4 medium, 2 small repos                                                                                                                                                                   & \begin{tabular}[c]{@{}l@{}}- Balanced Representation:\\ \hspace*{1em}Mitigate overrepresentation. \\ - Quality:\\ \hspace*{1em}Prioritize maturity, documentation, \\ \hspace*{1em}commit history.\end{tabular} & \begin{tabular}[c]{@{}l@{}}Stratified sampling. \\Manual curation. \end{tabular}              & 10 repos                                                             \\ \hline
        \end{tabular}%
        }
    \end{table*}

        A commit history scan then searched for terms such as "refactor," "rewrite," "enhance," and "optimize," indicating meaningful maintenance activity. Repositories also needed basic documentation with keywords like "usage," "install," or "example." This reduced the pool to 112 projects. These were classified by size using LOC: small (<500), medium (500–2000), and large (>2000). Table \ref{tab:osp_proj} lists their key attributes: repository popularity, development activity, contributor count, project age, and forks. A stratified sample of 10 projects (4 large, 4 medium, 2 small) was selected for evaluation.

    \item \textbf{Refactoring Codebases:} 
    The selected codebases in Table \ref{tab:osp_proj} were processed through the multi-agent system in a controlled environment. 

\end{itemize}
    
    \begin{table}[]
        \caption{List of Haskell-based OSP evaluated in this study}
        \label{tab:osp_proj}
        \centering
        \renewcommand{\arraystretch}{1}
        
        \begin{tabular}{|l@{\hspace{4pt}}|r@{\hspace{4pt}}|r@{\hspace{4pt}}|r@{\hspace{4pt}}|r@{\hspace{4pt}}|r@{\hspace{4pt}}|}
        \hline
        \textbf{Project} & \textbf{Stars}  & \textbf{Commits} & \textbf{Contributers} & \textbf{Age (years)} & \textbf{Forks} \\ \hline
        Cryptol           & 1,200           & 4,521   & 58                    & 11                   & 126            \\
        Granule           & 617             & 3,879   & 18                    & 8                    & 38             \\
        Obelisk           & 1,000           & 2,847   & 51                    & 6.9         & 107            \\
        Fused Effects     & 656             & 5,206            & 28                    & 6.6         & 53             \\
        Nix-tree          & 854             & 230     & 12                    & 4.8         & 17             \\
        Erd               & 1,830  & 152     & 24                    & 11.5                 & 153            \\
        Tetris            & 936    & 124     & 4                     & 7.9                  & 41             \\
        Greenclip         & 1,504           & 168              & 14                    & 8.3                  & 35             \\
        Xdg-inja          & 2,816           & 534              & 170                   & 3                    & 157            \\
        Bench             & 887             & 71               & 7                     & 9.1                  & 21             \\ \hline
        \textbf{Total}    & \textbf{12,300} & \textbf{17,732}  & \textbf{386}          & \textbf{}            & \textbf{748} \\ \hline
        \end{tabular}
        
    \end{table}
The codebases evaluation process involves three critical steps: deciding the evaluation criteria, selecting representative codebases, and executing the refactoring process while measuring performance.

\subsection{Evaluation}
Multi-agent system is evaluated using metrics specific to functional programming and software engineering. This ensures refactoring outputs meet established standards for quality, performance, and maintainability.
\begin{itemize}
  \item \textbf{Code Complexity Reduction:} Cyclomatic complexity and structural dependencies are measured pre-refactoring and post-refactoring using established software metrics. This provides a quantitative measure of simplification. To measure the complexity of the codebases, we utilized McCabe's cyclomatic complexity metric~\cite{mccabe1976complexity}, which quantifies the number of linearly independent paths through a program. Cyclomatic Complexity values were calculated using standard static analysis tools before and after the refactoring process. Higher values indicate increased complexity, while reductions suggest improved maintainability and readability of the code. The effectiveness of complexity reduction was benchmarked against prior studies. It indicates the program's complexity and maintainability. The formula for calculating CC is:
    \begin{equation}
        CC = E - N + 2P
    \end{equation}
    where \(E\): Number of edges in the control flow graph, \(N\): Number of nodes in the control flow graph, and \(P\): Number of connected components (typically 1 for a single program).
    
    For multiple functions, the total complexity is:
    \begin{equation}
         CC_{\text{Total}} = \sum_{i=1}^{n} CC_i
    \end{equation}
    High CC values indicate code that is harder to understand, test, and maintain. Code clarity and efficiency are improved by reducing CC through refactoring. 
    
    \item \textbf{Performance Benchmarks:} Runtime and memory usage are evaluated using GHC profiling tools. This aligns with prior studies on performance optimization in Haskell~\cite{gill2009worker}. Runtime refers to the execution time of a program, often measured in ticks or seconds. Memory usage refers to the total bytes allocated during execution. These metrics are critical for assessing the efficiency of a program. Using the Glasgow Haskell Compiler (GHC) profiling tools (+RTS options), the runtime and memory performance of the codebases were evaluated.

    \item \textbf{Comparison of Code Quality with HLint:} HLint is a static code analysis tool for Haskell that suggests stylistic improvements to enhance code readability and maintainability. HLint was used to evaluate these improvements in the refactored codebases.
\end{itemize}

This evaluation integrates automated testing frameworks \cite{prechelt2000empirical} and uses previous work on program synthesis and optimization in functional programming~\cite{chen2021evaluating}. These benchmarks provide validation for the multi-agent system's outputs.

\section{Results} \label{sec:results}

To assess the effectiveness of the proposed multi-agent system in refactoring Haskell code, we evaluated its performance on OSP. Evaluation focused on widely accepted software quality metrics. Each metric serving as an indicator of improved maintainability, performance, or code idiomaticity. As summarized in Table \ref{tab:post_refactor_improvments}, improvements were recorded across all metrics. We present a detailed analysis of the performance outcomes from applying the multi-agent refactoring system to the selected Haskell codebases. The evaluation framework was designed around OSP, allowing for an assessment of the system across organically developed software. The primary aim was to measure improvements across essential software quality metrics. We applied our multi-agent system to 10 open source projects. We measured the performance of the metrics before and after refactoring. Table \ref{tab:post_refactor_improvments} summarizes the percentage improvements for each group. Table \ref{tab:post_refactor_improvments} shows that our system reduced cyclomatic complexity by 8.90\%. It cut branching depth by 9.76\%. It eliminated 25.20\% of HLint suggestions. It removed 30.20\% of warnings and 11.97\% of errors. It improved runtime efficiency by 5.28\%. Finally, it reduced memory allocation by 21.27\%.

\begin{table*}[ht!]
\caption{Post-refactoring improvements in percentages (\%) on open source Haskell codebases}
\label{tab:post_refactor_improvments}
\centering
\renewcommand{\arraystretch}{1.9}
\resizebox{\textwidth}{!}{%
\begin{tabular}{|>{\hspace{4pt}}l>{\hspace{4pt}}r@{\hspace{4pt}}>{\hspace{4pt}}r@{\hspace{4pt}}>{\hspace{4pt}}r@{\hspace{4pt}}>{\hspace{4pt}}r@{\hspace{4pt}}>{\hspace{4pt}}r@{\hspace{4pt}}>{\hspace{4pt}}r@{\hspace{4pt}}>{\hspace{4pt}}r@{\hspace{4pt}}>{\hspace{4pt}}r@{\hspace{4pt}}|>{\hspace{4pt}}r@{\hspace{4pt}}|}
\hline
 \multicolumn{1}{|l|}{\textbf{}}                              & \multicolumn{3}{c|}{Code Complexity Improvement (\%)}                                                   & \multicolumn{3}{c|}{Code Quality Improvement (\%)}                              & \multicolumn{2}{c|}{Code Performance Improvement (\%)}    \\
\multicolumn{1}{|l|}{\textbf{Project}}              & \textbf{Lines of Code} & \textbf{Cyclomatic Complexity} & \multicolumn{1}{r|}{\textbf{Branching Depth}} & \textbf{Suggestions} & \textbf{Warnings} & \multicolumn{1}{r|}{\textbf{Errors}} & \textbf{Runtime Efficiency} & \textbf{Memory Allocation} \\ \hline
\multicolumn{1}{|l|}{Cryptol}                       & 5.88                   & 4.28                  & \multicolumn{1}{r|}{8.63}                     & 11.12                & 1.15              & \multicolumn{1}{r|}{5.32}            & 8.03                        & 38.7                       \\
\multicolumn{1}{|l|}{Granule}                       & 8.09                   & 17.99                 & \multicolumn{1}{r|}{6.57}                     & 22.76                & 1.04              & \multicolumn{1}{r|}{4.24}            & 3.35                        & 18.9                       \\
\multicolumn{1}{|l|}{Obelisk}                       & 15.36                  & 0.82                  & \multicolumn{1}{r|}{9.38}                     & 4.35                 & 5.40              & \multicolumn{1}{r|}{2.05}            & 11.50                       & 19.4                       \\
\multicolumn{1}{|l|}{Fused Effects}                 & 1.09                   & 3.49                  & \multicolumn{1}{r|}{14.29}                    & 31.60                & 37.5              & \multicolumn{1}{r|}{2.10}            & 7.15                        & 18.1                       \\ \cline{1-9} 
\textbf{Average (Large)}                           & \textbf{7.61}          & \textbf{6.65}         & \textbf{9.72}                                 & \textbf{17.46}       & \textbf{11.27}    & \textbf{3.43}                        & \textbf{7.51}               & \textbf{23.78}             \\ \cline{1-9} 
\multicolumn{1}{|l|}{Nix-tree}                      & 7.16                   & 3.02                  & \multicolumn{1}{r|}{0}                        & 25.00                & 57.14             & \multicolumn{1}{r|}{5.41}            & 10.76                       & 67.7                       \\
\multicolumn{1}{|l|}{Erd}                           & 15.74                  & 1.98                  & \multicolumn{1}{r|}{12.5}                     & 0                    & 0                 & \multicolumn{1}{r|}{6.34}            & 3.67                        & 47.1                       \\
\multicolumn{1}{|l|}{Tetris}                        & 0.43                   & 22.14                 & \multicolumn{1}{r|}{0}                        & 40.00                & 21.20             & \multicolumn{1}{r|}{2.56}            & 8.34                        & 0.4                        \\
\multicolumn{1}{|l|}{Greenclip}                     & 0.95                   & 17.98                 & \multicolumn{1}{r|}{0}                        & 60.00                & 100.00            & \multicolumn{1}{r|}{59.38}           & 0                           & 0                          \\ \cline{1-9} 
\textbf{Average (Medium)}                          & \textbf{6.07}          & \textbf{11.28}        & \textbf{3.13}                                 & \textbf{31.25}       & \textbf{44.59}    & \textbf{18.42}                       & \textbf{5.69}               & \textbf{28.8}              \\ \cline{1-9} 
\multicolumn{1}{|l|}{Xdg-inja}                      & 30.72                  & 17.59                 & \multicolumn{1}{r|}{42.85}                    & 57.14                & 78.57             & \multicolumn{1}{r|}{32.30}           & 0                           & 0.6                        \\
\multicolumn{1}{|l|}{Bench}                         & 3.54                   & 8.33                  & \multicolumn{1}{r|}{50.00}                    & 0                    & 0                 & \multicolumn{1}{r|}{0}               & 0                           & 1.8                        \\ \cline{1-9} 
\textbf{Average (Small)}                           & \textbf{17.13}         & \textbf{12.96}        & \textbf{46.42}                                & \textbf{28.57}       & \textbf{39.28}    & \textbf{16.15}                       & \textbf{0}                  & \textbf{1.2}               \\ \cline{1-9} 
\textbf{Total Average}                             & \textbf{8.90}          & \textbf{9.76}         & \textbf{14.42}                                & \textbf{25.20}       & \textbf{30.2}     & \textbf{11.97}                       & \textbf{5.28}               & \textbf{21.27}             \\ \hline
\end{tabular}%
}
\end{table*}

\begin{figure*}[h]
    \centering
    
    \includegraphics[width=0.99\textwidth, clip, trim=50 120 200 80]{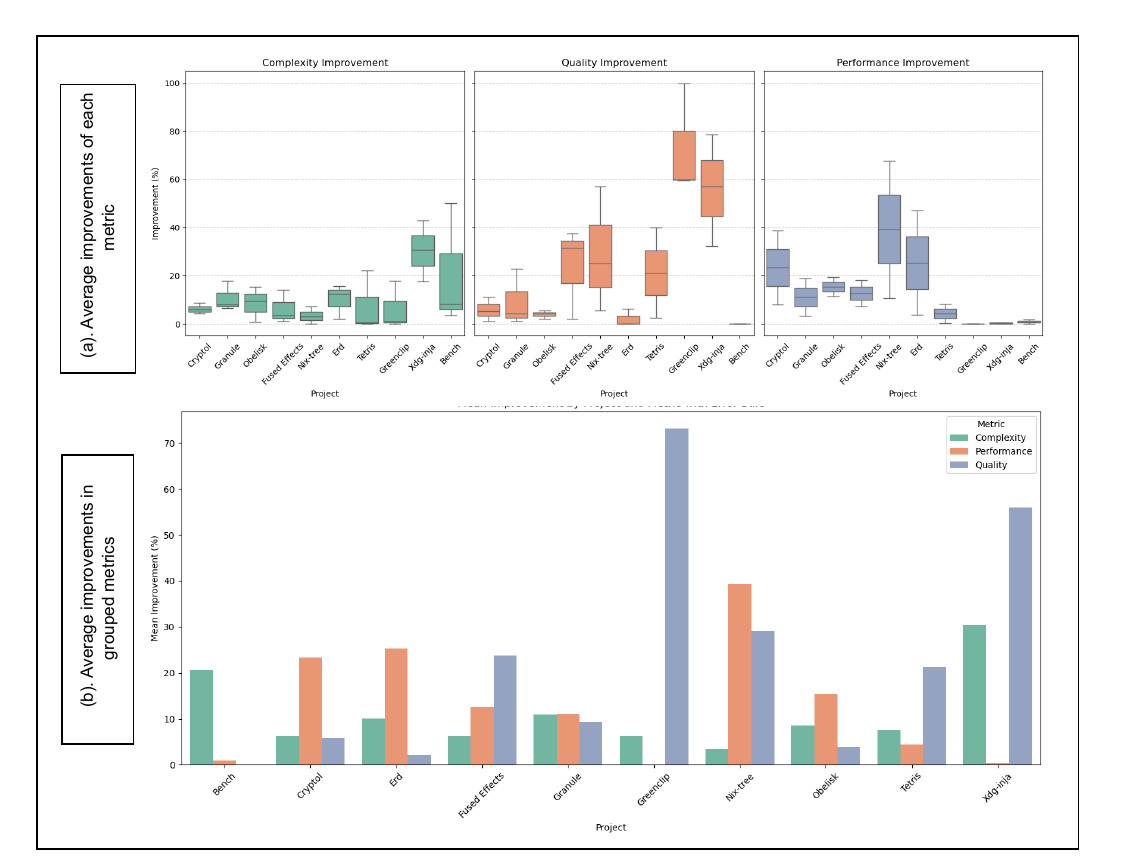}
    \caption{Overall OSP post-refactor improvements comparison of different metrics}
    \label{fig:Overall OSP Post-refactor improvements comparison of different metrics}
\end{figure*}

\subsection{RQ1: Effects of LLM-Based Multi-Agent Systems on Haskell Codebases: Cyclomatic Complexity, Runtime, and Memory Usage}
\textbf{Cyclomatic Complexity and Branching Depth Reduction}
Cyclomatic complexity, a foundational software metric, quantifies the number of linearly independent paths through a program’s source code. Reducing this metric generally correlates with enhanced maintainability, reduced cognitive load for developers, and lower likelihood of defects. Through the multi-agent system, significant reductions in this metric were achieved. An average reduction of 8.90\% was observed, demonstrating that the multi-agent system effectively simplified branching and conditional structures within organically written Haskell code. This suggests that the multi-agent system not only handles real-world complexities but also excels in applying targeted simplifications when the structural boundaries are more predictable.

System measures branching depth as the maximum number of nested control structures (e.g., if, case, and loops) within a function. Deep nesting increases cognitive load and raises the risk of bugs. Extracting complex case expressions into small helper functions by refactoring nested conditionals into flat guard clauses and adopting early exits to remove needless branches. Our agents cut the average branch depth by 14.42\%. This reduction translates to fewer indentation levels and a clearer logical path in each function. Shallower nesting improves readability, eases future modifications, and reduces error proneness in Haskell code.

These reductions are especially important in long-lived or collaborative codebases, where high complexity can slow development and increase maintenance risk.

\textbf{Runtime Efficiency Gains}: Runtime performance is crucial for any production-level software system. Enhancements in this area were primarily driven by the multi-agent system’s ability to refactor inefficient recursive patterns and eliminate superfluous computation. Among the OSP projects, runtime efficiency improved by an average of 5.28\%, reflecting the multi-agent system’s capacity to navigate and optimize complex, often unstructured logic. This showed that multi-agent system-generated transformations led to meaningful reductions in execution time.

\begin{figure*}[h]
    \centering
    
    \includegraphics[width=0.99\textwidth, clip, trim=50 460 50 100]{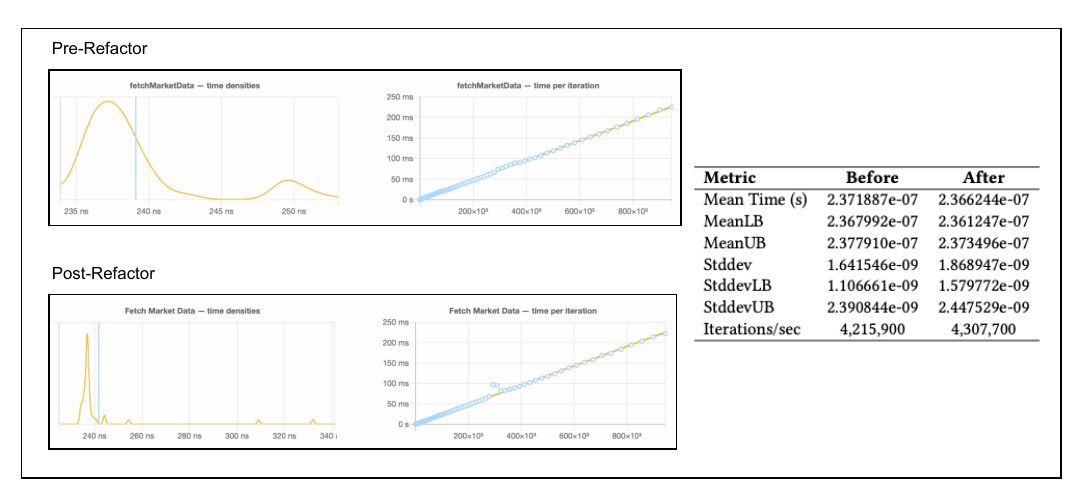}
    \caption{Pre-refactor and post-refactor performance comparison of function}
    \label{fig:Pre-refactor and Post-refactor Performance comparison of Function}
\end{figure*}

\textbf{Memory Allocation Optimization}: The multi-agent system also impacted memory efficiency, a critical factor for applications deployed in resource-constrained environments. Optimizations were achieved through a variety of strategies, including restructuring data flows, reducing intermediate data structures, and refining state-handling mechanisms. We measured execution time and memory allocation with GHC profiling. Fig. \ref{fig:Pre-refactor and Post-refactor Performance comparison of Function} shows a representative function’s runtime profile before and after refactoring. OSP codebases experienced an average memory usage reduction of 21.27\% and improved runtime efficiency by 5.28\%.These gains underline the multi-agent system’s competence in managing both abstract and concrete data manipulations with precision.

\begin{tcolorbox}[colback=gray!10, colframe=black, title=Key Takeaways from RQ1]

Our multi-agent system demonstrates effectiveness in reducing complexity, optimizing runtime efficiency, and  lowering memory usage, thereby enhancing overall maintainability and performance of Haskell codebases. The results indicate the potential of LLM-based multi-agent systems to support the refactoring of functional programming languages.

\end{tcolorbox}

\subsection{RQ2: Evaluating the Impact of Multi-agent Approaches on Refactoring Workflows in Functional Programming Languages}
\textbf{HLint-based Code Quality}:
To assess stylistic and semantic code quality improvements, the refactored code was subjected to HLint, a widely used static analysis tool in the Haskell ecosystem. Post-refactoring assessments revealed that the multi-agent system had successfully removed redundant expressions, enforced consistent naming conventions, and enhanced code readability. OSP projects showed improvements in their HLint scores, supporting the claim that the multi-agent system promotes cleaner and more maintainable Haskell code. This also indirectly validates the multi-agent system’s semantic understanding of the code, as many HLint suggestions involve deep structural and contextual insight.

Our multi-agent system targets HLint flags and compiler diagnostics with precise refactorings—adding missing type signatures, pruning unused bindings, swapping partial functions for total ones, and streamlining expressions. HLint suggestions fell by 25.20\%, compiler warnings by 30.20\%, and compiler errors by 11.97\%. These gains confirm that our agents enforce idiomatic Haskell, eradicate common code smells, and resolve release-blocking issues.

\begin{figure}[h]
\centering
\includegraphics[scale=1.0, clip, trim=15 450 250 10]{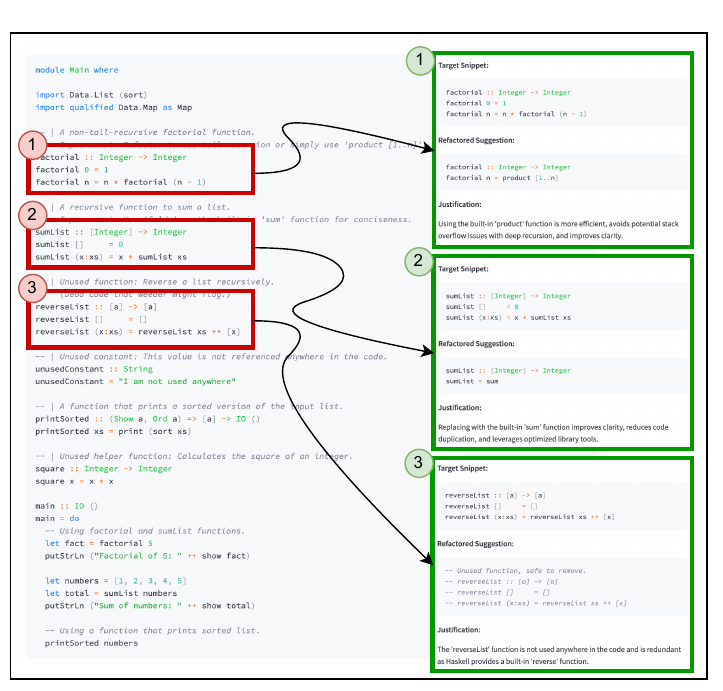}
\caption{Common refactoring patterns (code smells)}
\label{fig:Refactoring_Suggestions}
\end{figure}

\textbf{Common Refactoring Patterns}:
Fig. \ref{fig:Refactoring_Suggestions} illustrates three frequent code-smell corrections made by our agents: replacing nested map/filter chains with compositions, using operator sections instead of lambdas, and renaming functions for clarity.

\textbf{Size Category Breakdown}:
We further analyzed results by project size. Table 3 reports average complexity, quality, and performance improvements for large, medium, and small projects.

\begin{table}[ht]
\caption{Category-wise comparison of Haskell projects (OSP)}
\label{tab:category_analysis}
\centering
\renewcommand{\arraystretch}{1.5}

\begin{tabular}{|>{\hspace{6pt}}l@{\hspace{6pt}}|>{\hspace{6pt}}c@{\hspace{6pt}}|>{\hspace{6pt}}c@{\hspace{6pt}}|>{\hspace{6pt}}c@{\hspace{6pt}}|}
\hline
\textbf{Metric} & \textbf{Large} & \textbf{Medium} & \textbf{Small} \\ \hline
\textbf{Avg. Code Complexity (\%)}    & 7.99                            & 6.83                                & 25.52                                 \\
\textbf{Avg. Code Quality (\%)}   & 10.72                            & 31.42                                & 28                                 \\
\textbf{Avg. Code Performance (\%)}    & 15.65                            & 17.24                                & 0.6                                 \\ \hline
\end{tabular}
\end{table}

Table \ref{tab:category_analysis} confirms that small projects saw the largest complexity drop (25.52\%), medium projects gained most in HLint quality (31.42\%), and large projects achieved consistent, if more modest, gains across all metrics

\begin{tcolorbox}[colback=gray!10, colframe=black, title=Key Takeaways from RQ2]

The use of specialized agents improves the code quality and adaptability of the refactoring workflow. The multi-agent collaboration handles complex code transformations, improving both code quality and semantic correctness, making this approach highly beneficial for functional programming languages like Haskell.

\end{tcolorbox}

\subsection{Consolidated Observations}
The results demonstrate that the LLM-based multi-agent system provides reliable and scalable improvements in software quality metrics. Consistently strong performance on real-world OSP codebases confirms the system’s adaptability to codebases shaped by diverse development practices, histories, and conventions. Collectively, these findings affirm the multi-agent system as a generalizable solution for modern code refactoring challenges in functional programming.

To further illustrate these improvements, we provide a visual summary of the average improvements in individual metrics and grouped metrics (see Fig. \ref{fig:Overall OSP Post-refactor improvements comparison of different metrics}). Part (a) of the figure highlights gains across metrics. Part (b) emphasizes grouped metric improvements, showing that complexity and quality-related metrics saw notable enhancements, underscoring the overall effectiveness the multi-agent refactoring approach. These visuals corroborate our quantitative findings and further validate the capability of the multi-agent system to deliver impactful, practical improvements in functional programming codebases.

\section{Discussion} 
\label{sec:discussion}

This section examines the findings in relation to the research questions and the broader objectives of automated software engineering. The observed improvements across performance and quality metrics support the hypothesis that a LLM-based multi-agent system can refactor functional programming code effectively. In addition, the results suggest potential implications for the development of future tools, design patterns, and practices in software maintenance.

\subsection{LLM-based Refactoring Efficiency (RQ1)}

Our study found that the proposed multi-agent system reduced cyclomatic complexity by an average of \textbf{8.90\%}, improved runtime efficiency by \textbf{5.28\%}, and optimized memory allocation by \textbf{21.27\%}. These findings indicate that the system was able to simplify, making the code more readable and maintainable. Furthermore, the multi-agent system demonstrated performance in optimizing memory usage, which is helpful for resource-constrained environments.These results align with existing study on refactoring and code optimization. Similar studies have shown that tools can improve cyclomatic complexity and runtime efficiency in imperative languages, such as those reported by Mens and Tourwé \cite{mens2004survey}. However, our findings are particularly for functional programming languages like Haskell, which involve unique constructs such as monads and lazy evaluation, making traditional refactoring tools less effective. \textbf{Implications:} For researchers, this study provides a proof of concept for applying multi-agent systems in functional programming code refactoring, suggesting that these systems can be adapted to handle domain-specific challenges in functional languages \cite{thompson2013refactoring}. For practitioners, the findings indicate that LLM-based multi-agent systems can automate refactoring tasks \cite{nascimento2023self}, leading to more maintainable, optimized code with minimal human intervention. Humans in the loop are still needed because the refactored code conflicts with organizational policies or regulatory standards.

\subsection{Multi-Agent-Based Refactoring Impact (RQ2)}

The multi-agent approach improved the refactoring workflow, leading to a \textbf{30.20\%} reduction in HLint warnings and \textbf{11.97\%} reduction in errors. This suggests that the multi-agent system was effective in enhancing code quality by addressing both stylistic and semantic issues. The iterative collaboration between specialized agents allowed for a flexible workflow that could adapt to different challenges presented by the codebase.Our results align with previous work on multi-agent systems in software engineering, such as AyshwaryaLakshmi \textit{et al}. \cite{ayshwaryalakshmi2013agent}, who demonstrated that multi-agent systems can improve the efficiency of distributed refactoring tasks. Moreover, our findings build on the work by Svyatkovskiy \textit{et al}. \cite{svyatkovskiy2020intellicode}, who explored the role of AI in assisting code generation and refactoring. However, unlike earlier studies, our work extends these concepts into functional programming, an area with unique refactoring challenges. \textbf{Implications:} The significant reductions highlight the practical benefits of multi-agent systems for improving code quality. For researchers, these results suggest that multi-agent systems can provide  improvements in code quality within functional programming contexts \cite{gyori2013crossing}, such as Haskell, which pose unique challenges. For practitioners, the findings suggest that incorporating multi-agent-based tools can directly enhance the efficiency of the refactoring process by addressing both stylistic and semantic issues \cite{tan2024fixing}, ultimately contributing to software maintenance practices \cite{alomar2021refactoring}.

\subsection{Limitations}

There are certain limitations to this work. First limitation is that we evaluate the performance of the multi-agent system with GPT-4o, without comparing the results to other LLMs or existing refactoring tools. As the system is model-agnostic, using different models could lead to variations in results. The focus is also restricted to Haskell, a functional programming language. Future work should investigate whether the multi-agent system can be extended to imperative or object-oriented languages. The system may also produce refactorings that conflict with organizational policies or regulatory standards, particularly in fields with strict compliance requirements. Furthermore, the limited token window of large language models results in the loss of context over long code sequences. This can lead to outputs that are inconsistent or incorrect in large-scale (LOC in millions) projects. There are also challenges regarding accountability for AI-generated code. Issues related to ownership, liability, and adherence to regulatory standards need to be addressed. Additionally, tracing errors in multi-agent systems remains difficult, as outputs are contingent upon interactions between multiple agents, complicating the identification and resolution of issues. 

\section{Threats to Validity} \label{sec:threat}

\noindent\par\textbf{Internal validity} concerns the accuracy of the experimental design and the correctness of our measurements. Controlled benchmarking tools were employed to evaluate the observed improvements in complexity and efficiency. However, variations in initial code quality, developer implementation choices, and underlying system configurations may introduce biases. Furthermore, the automated refactoring process could have inadvertently altered code structures in ways that were not fully captured by our metrics.

\noindent\par\textbf{External validity} addresses how well our findings can be applied to broader scenarios. The evaluation was conducted on a limited number of Haskell codebases, primarily from open-source repositories, which may not fully represent the diversity of functional programming applications in industry or academia. The results may vary for significantly larger, highly optimized, or domain-specific Haskell projects. To enhance the applicability of our results, future work should examine a more extensive set of codebases across various fields and levels of complexity.

\noindent\par\textbf{Construct validity} examines whether the selected metrics accurately reflect the efficacy of our approach. Although quantitative measures like cyclomatic complexity, runtime efficiency, and memory allocation offer valuable insights, we do not fully capture qualitative aspects such as developer experience. Subjective elements, including how programmers view the legibility of refactored code or the simplicity of future updates, are not addressed. Incorporating qualitative assessments from experienced Haskell developers would strengthen our evaluation.

\section{Conclusions and Future Research} \label{sec:conclusions}

We presented an LLM-based multi-agent system for the intelligent refactoring of Haskell codebases. Through a structured system comprising agents for analysis, transformation, and validation, the system achieves automated, context-sensitive improvements in code structure and quality. The multi-agent system was tested on open source projects (OSP). The results showed improvements in all the dimensions measured, including cyclomatic complexity, runtime performance, memory usage, and quality based on HLint. These outcomes validate the multi-agent system’s capability to support high-impact code maintenance while minimizing manual intervention. Notably, the system demonstrated reliable adaptability for real-world and organically developed software. In summary, the proposed multi-agent system exemplifies the convergence of LLM and multi-agent coordination for scalable, intelligent code refactoring. We plan to expand the system to include additional programming paradigms and explore the integration of learning agents capable of evolving based on transformation feedback and historical performance data.

Future research should focus on the improvments of context management and long-term memory to overcome the token window limits. This will help agents to process longer code without context loss. Models for human-AI collaboration and models for  adaptive autonomy should be explored to improve system interaction. These models can help the system learn from past refactorings and adjust strategies based on feedback. In addition, frameworks for safety, security, and compliance are required to ensure that the suggested refactorings meet the required compliance with organizational and regulatory standards. Finally, the research area of error traceability in multi-agent systems prompted by development should be broadened in scope. Developing mechanisms to track error origins in agent interactions will help in debugging and fault detection.

%
%
%
%

\bibliographystyle{splncs04}
\bibliography{references}

\begin{thebibliography}{10}
\providecommand{\url}[1]{\texttt{#1}}
\providecommand{\urlprefix}{URL }
\providecommand{\doi}[1]{https://doi.org/#1}

\bibitem{abdallah2011dynamic}
Abdallah, C., Bouziane, H.: Dynamic maintenance and evolution of critical components-based software using multi agent systems. Computer and Information Science  \textbf{4}(5), ~78 (2011)

\bibitem{allamanis2018survey}
Allamanis, M., Barr, E.T., Devanbu, P., Sutton, C.: A survey of machine learning for big code and naturalness. ACM Computing Surveys (CSUR)  \textbf{51}(4),  1--37 (2018)

\bibitem{alomar2021refactoring}
AlOmar, E.A., AlRubaye, H., Mkaouer, M.W., Ouni, A., Kessentini, M.: Refactoring practices in the context of modern code review: An industrial case study at xerox. In: 2021 IEEE/ACM 43rd International Conference on Software Engineering: Software Engineering in Practice (ICSE-SEIP). pp. 348--357. IEEE (2021)

\bibitem{ayshwaryalakshmi2013agent}
AyshwaryaLakshmi, S., Mary, S.S.A., Vadivu, S.S.: Agent based tool for topologically sorting badsmells and refactoring by analyzing complexities in source code. In: 2013 Fourth International Conference on Computing, Communications and Networking Technologies (ICCCNT). pp.~1--7. IEEE (2013)

\bibitem{baumgartner2024ai}
Baumgartner, N., Iyenghar, P., Schoemaker, T., Pulverm{\"u}ller, E.: Ai-driven refactoring: A pipeline for identifying and correcting data clumps in git repositories. Electronics  \textbf{13}(9), ~1644 (2024)

\bibitem{bragilevsky2021haskell}
Bragilevsky, V.: Haskell in Depth. Simon and Schuster (2021)

\bibitem{brown2011expression}
Brown, C., Li, H., Thompson, S.: An expression processor: a case study in refactoring haskell programs. In: Trends in Functional Programming: 11th International Symposium, TFP 2010, Norman, OK, USA, May 17-19, 2010. Revised Selected Papers 11. pp. 31--49. Springer (2011)

\bibitem{brown2020language}
Brown, T., Mann, B., Ryder, N., Subbiah, M., Kaplan, J.D., Dhariwal, P., Neelakantan, A., Shyam, P., Sastry, G., Askell, A., et~al.: Language models are few-shot learners. Advances in neural information processing systems  \textbf{33},  1877--1901 (2020)

\bibitem{chen2021evaluating}
Chen, M., Tworek, J., Jun, H., Yuan, Q., Pinto, H.P.D.O., Kaplan, J., Edwards, H., Burda, Y., Joseph, N., Brockman, G., et~al.: Evaluating large language models trained on code. arXiv preprint arXiv:2107.03374  (2021)

\bibitem{cheng2024exploring}
Cheng, Y., Zhang, C., Zhang, Z., Meng, X., Hong, S., Li, W., Wang, Z., Wang, Z., Yin, F., Zhao, J., et~al.: Exploring large language model based intelligent agents: Definitions, methods, and prospects. arXiv preprint arXiv:2401.03428  (2024)

\bibitem{feng2024genetic}
Feng, C., Sun, Y., Li, K., Zhou, P., Lv, J., Lu, A.: Genetic auto-prompt learning for pre-trained code intelligence language models. arXiv preprint arXiv:2403.13588  (2024)

\bibitem{figueroa2021monads}
Figueroa, I., Leger, P., Fukuda, H.: Which monads haskell developers use: An exploratory study. Science of Computer Programming  \textbf{201},  102523 (2021)

\bibitem{gill2009worker}
Gill, A., Hutton, G.: The worker/wrapper transformation. Journal of Functional Programming  \textbf{19}(2),  227--251 (2009)

\bibitem{griffin2021hexagonal}
Griffin, J., Griffin, J.: Hexagonal-driven development. Domain-Driven Laravel: Learn to Implement Domain-Driven Design Using Laravel pp. 521--544 (2021)

\bibitem{guo2024large}
Guo, T., Chen, X., Wang, Y., Chang, R., Pei, S., Chawla, N.V., Wiest, O., Zhang, X.: Large language model based multi-agents: A survey of progress and challenges. arXiv preprint arXiv:2402.01680  (2024)

\bibitem{gyori2013crossing}
Gyori, A., Franklin, L., Dig, D., Lahoda, J.: Crossing the gap from imperative to functional programming through refactoring. In: Proceedings of the 2013 9th Joint Meeting on Foundations of Software Engineering. pp. 543--553 (2013)

\bibitem{he2005hexagonal}
He, X., Jia, W.: Hexagonal structure for intelligent vision. In: 2005 International conference on information and communication technologies. pp. 52--64. IEEE (2005)

\bibitem{hu2015functional}
Hu, Z., Hughes, J., Wang, M.: How functional programming mattered. National Science Review  \textbf{2}(3),  349--370 (2015)

\bibitem{hua2023war}
Hua, W., Fan, L., Li, L., Mei, K., Ji, J., Ge, Y., Hemphill, L., Zhang, Y.: War and peace (waragent): Large language model-based multi-agent simulation of world wars. arXiv preprint arXiv:2311.17227  (2023)

\bibitem{huang2024levels}
Huang, Y.: Levels of ai agents: from rules to large language models. arXiv preprint arXiv:2405.06643  (2024)

\bibitem{hudak1992gentle}
Hudak, P., Fasel, J.H.: A gentle introduction to haskell. ACM Sigplan Notices  \textbf{27}(5),  1--52 (1992)

\bibitem{ishizue2024improved}
Ishizue, R., Sakamoto, K., Washizaki, H., Fukazawa, Y.: Improved program repair methods using refactoring with gpt models. In: Proceedings of the 55th ACM Technical Symposium on Computer Science Education V. 1. pp. 569--575 (2024)

\bibitem{jiang2024survey}
Jiang, J., Wang, F., Shen, J., Kim, S., Kim, S.: A survey on large language models for code generation. arXiv preprint arXiv:2406.00515  (2024)

\bibitem{khalil2016onion}
Khalil, M.E., Ghani, K., Khalil, W.: Onion architecture: a new approach for xaas (every-thing-as-a service) based virtual collaborations. In: 2016 13th Learning and Technology Conference (L\&T). pp.~1--7. IEEE (2016)

\bibitem{li2005haskell}
Li, H., Thompson, S., Reinke, C.: The haskell refactorer, hare, and its api. Electronic Notes in Theoretical Computer Science  \textbf{141}(4),  29--34 (2005)

\bibitem{mccabe1976complexity}
McCabe, T.J.: A complexity measure. IEEE Transactions on software Engineering (4),  308--320 (1976)

\bibitem{mens2004survey}
Mens, T., Tourw{\'e}, T.: A survey of software refactoring. IEEE Transactions on software engineering  \textbf{30}(2),  126--139 (2004)

\bibitem{nascimento2023self}
Nascimento, N., Alencar, P., Cowan, D.: Self-adaptive large language model (llm)-based multiagent systems. In: 2023 IEEE International Conference on Autonomic Computing and Self-Organizing Systems Companion (ACSOS-C). pp. 104--109. IEEE (2023)

\bibitem{orchard2014embedding}
Orchard, D., Petricek, T.: Embedding effect systems in haskell. In: Proceedings of the 2014 ACM SIGPLAN Symposium on Haskell. pp. 13--24 (2014)

\bibitem{peyton1993imperative}
Peyton~Jones, S.L., Wadler, P.: Imperative functional programming. In: Proceedings of the 20th ACM SIGPLAN-SIGACT symposium on Principles of programming languages. pp. 71--84 (1993)

\bibitem{pomian2024next}
Pomian, D., Bellur, A., Dilhara, M., Kurbatova, Z., Bogomolov, E., Bryksin, T., Dig, D.: Next-generation refactoring: Combining llm insights and ide capabilities for extract method. In: 2024 IEEE International Conference on Software Maintenance and Evolution (ICSME). pp. 275--287. IEEE (2024)

\bibitem{prechelt2000empirical}
Prechelt, L.: An empirical comparison of seven programming languages. Computer  \textbf{33}(10),  23--29 (2000)

\bibitem{rich1985layered}
Rich, C.: The layered architecture of a system for reasoning about programs. In: IJCAI. vol.~9, pp. 540--546. Citeseer (1985)

\bibitem{dos2015autorefactoring}
dos Santos~Neto, B.F., Ribeiro, M., Da~Silva, V.T., Braga, C., De~Lucena, C.J.P., de~Barros~Costa, E.: Autorefactoring: A platform to build refactoring agents. Expert systems with applications  \textbf{42}(3),  1652--1664 (2015)

\bibitem{shirafuji2023refactoring}
Shirafuji, A., Oda, Y., Suzuki, J., Morishita, M., Watanobe, Y.: Refactoring programs using large language models with few-shot examples. In: 2023 30th Asia-Pacific Software Engineering Conference (APSEC). pp. 151--160. IEEE (2023)

\bibitem{ShahbazMaSHaskellGitHub2025}
Siddeeq, S.: Intelligent haskell code refactoring using mulit-agent system (2025), \url{https://github.com/GPT-Laboratory/Intelligent-Haskell-Code-Refactoring}, accessed: 31-Jan-2025

\bibitem{svyatkovskiy2020intellicode}
Svyatkovskiy, A., Deng, S.K., Fu, S., Sundaresan, N.: Intellicode compose: Code generation using transformer. In: Proceedings of the 28th ACM joint meeting on European software engineering conference and symposium on the foundations of software engineering. pp. 1433--1443 (2020)

\bibitem{tan2024fixing}
Tan, I., Poskitt, C.M.: Fixing your own smells: Adding a mistake-based familiarisation step when teaching code refactoring. In: Proceedings of the 55th ACM Technical Symposium on Computer Science Education V. 1. pp. 1307--1313 (2024)

\bibitem{thompson2013refactoring}
Thompson, S., Li, H.: Refactoring tools for functional languages. Journal of Functional Programming  \textbf{23}(3),  293--350 (2013)

\bibitem{wadler1992essence}
Wadler, P.: The essence of functional programming. In: Proceedings of the 19th ACM SIGPLAN-SIGACT symposium on Principles of programming languages. pp. 1--14 (1992)

\bibitem{white2024chatgpt}
White, J., Hays, S., Fu, Q., Spencer-Smith, J., Schmidt, D.C.: Chatgpt prompt patterns for improving code quality, refactoring, requirements elicitation, and software design. In: Generative AI for Effective Software Development, pp. 71--108. Springer (2024)

\bibitem{wooldridge1995intelligent}
Wooldridge, M., Jennings, N.R.: Intelligent agents: Theory and practice. The knowledge engineering review  \textbf{10}(2),  115--152 (1995)

\end{thebibliography}




\end{document}